# Deterministic Automata
# for the (F,G)-fragment of LTL

Jan Křetínský[1,2][*] and Javier Esparza[1]

[1] Fakultät für Informatik, Technische Universität München, Germany
{jan.kretinsky,esparza}@in.tum.de
[2] Faculty of Informatics, Masaryk University, Brno, Czech Republic

**Abstract.** When dealing with linear temporal logic properties in the setting of e.g. games or probabilistic systems, one often needs to express them as deterministic omega-automata. In order to translate LTL to deterministic omega-automata, the traditional approach first translates the formula to a non-deterministic Büchi automaton. Then a determinization procedure such as of Safra is performed yielding a deterministic $\omega$-automaton. We present a direct translation of the (**F**,**G**)-fragment of LTL into deterministic $\omega$-automata with no determinization procedure involved. Since our approach is tailored to LTL, we often avoid the typically unnecessarily large blowup caused by general determinization algorithms. We investigate the complexity of this translation and provide experimental results and compare them to the traditional method.

## 1 Introduction

The $\omega$-regular languages play a crucial role in formal verification of linear time properties, both from a theoretical and a practical point of view. For model-checking purposes one can comfortably represent them using nondeterministic Büchi automata (NBW), since one only needs to check emptiness of the intersection of two NBWs corresponding to the system and the negation of the property, and NBWs are closed under intersection. However, two increasingly important problems require to represent $\omega$-regular languages by means of *deterministic* automata. The first one is synthesis of reactive modules for LTL specifications, which was theoretically solved by Pnueli and Rosner more than 20 years ago [PR88], but is recently receiving a lot of attention (see the references below). The second one is model checking Markov decision processes (see e.g. [BK08]), where impressive advances in algorithmic development and tool support are quickly extending the range of applications.

It is well known that NBWs are strictly more expressive then their deterministic counterpart, and so cannot be determinized. The standard theoretical solution to this problem is to translate NBW into deterministic Rabin automata (DRW) using Safra's construction [Saf88] or a recent improvement by Piterman [Pit06]. However, it is commonly accepted that Safra's construction is difficult to handle algorithmically due to its "messy state space" [Kup12]. Many possible strategies for solving this problem have been investigated. A first one is to avoid Safra's construction altogether. A Safraless approach that reduces the synthesis problem to emptiness of nondeterministic Büchi tree automata has been proposed in [KV05,KPV06]. The approach has had considerable success, and has been implemented in [JB06]. Another strategy is to use heuristics to improve Safra's construction, a path that has been followed in [KB06,KB07] and has produced the ltl2dstar tool [Kle]. Finally, a third strategy is to search for more efficient or simpler algorithms for subclasses of $\omega$-regular languages. A natural choice is to investigate classes of LTL formulas. While LTL is not as expressive as NBW, the complexity of the translation of LTL to DRW still has $2^{2^{\Theta(n)}}$ complexity [KR10]. However, the structure of NBWs for LTL formulas can be exploited to construct a symbolic description of a deterministic parity automaton [MS08]. Fragments of LTL have also been studied. In [AT04], single exponential translations for some simple fragments are presented. Piterman et al. propose in [PPS06] a construction for reactivity(1) formulas that produces in cubic time a symbolic representation of the automaton. The construction has been implemented in the ANZU tool [JGWB07].

---

[*] The author is a holder of Brno PhD Talent Financial Aid and is supported by the Czech Science Foundation, grant No. P202/12/G061.

Despite this impressive body of work, the problem cannot yet be considered solved. This is particularly so for applications to probabilistic model checking. Since probabilistic model checkers need to deal with linear arithmetic, they profit much less from sophisticated symbolic representations like those used in [PPS06,MS08], or from the Safraless approach which requires to use tree automata. In fact, to the best of our knowledge no work has been done so far in this direction. The most successful approach so far is the one followed by the ltl2dstar tool, which explicitly constructs a reduced DRW. In particular, the ltl2dstar has been reimplemented in PRISM [KNP11], the leading probabilistic model checker.

However, the work carried in [KB06,KB07] has not considered the development of specific algorithms for fragments of LTL. This is the question we investigate in this paper: is it possible to improve on the results of ltl2dstar by restricting attention to a subset of LTL? We give an affirmative answer by providing a very simple construction for the (**F**,**G**)-fragment of LTL, i.e., the fragment generated by boolean operations and the temporal operators **F** and **G**. Our construction is still double exponential in the worst case, but is algorithmically very simple. We construct a deterministic Muller automaton for a formula $\varphi$ of the fragment with a very simple state space: boolean combinations of formulas of the closure of $\varphi$. This makes the construction very suitable for applying reductions based on logical equivalences: whenever some logical rule shows that two states are logically equivalent, they can be merged. (This fact is also crucial for the success of the constructions from LTL to NBW.) Since the number of Muller accepting sets can be very large, we also show that the Muller condition of our automata admits a compact representation as a generalized Rabin acceptance condition. We also show how to efficiently transform this automaton to a standard Rabin automaton. Finally, we report on an implementation of the construction, and present a comparison with ltl2dstar. We show that our construction leads to substantially smaller automata for formulas expressing typical fairness conditions, which play a very important rôle in probabilistic model checking. For instance, while ltl2dstar produces an automaton with over one million states for the formula $\bigwedge_{i=1}^{3}(\mathbf{GF}a_i \to \mathbf{GF}b_i)$, our construction delivers an automaton with 1560 states.

## 2  Linear Temporal Logic

This section recalls the notion of linear temporal logic (LTL) [Pnu77].

**Definition 1 (LTL Syntax).** *The formulae of the ($\mathbf{F}$,$\mathbf{G}$)-fragment of linear temporal logic are given by the following syntax:*

$$\varphi ::= a \mid \neg a \mid \varphi \wedge \varphi \mid \varphi \vee \varphi \mid \mathbf{F}\varphi \mid \mathbf{G}\varphi$$

*where $a$ ranges over a finite fixed set $Ap$ of atomic propositions.*

We use the standard abbreviations $\mathbf{tt} := a \vee \neg a$, $\mathbf{ff} ;= a \wedge \neg a$. We only have negations of atomic propositions, as negations can be pushed inside due to the equivalence of $\mathbf{F}\varphi$ and $\neg \mathbf{G} \neg \varphi$.

**Definition 2 (LTL Semantics).** *Let $w \in (2^{Ap})^\omega$ be a word. The ith letter of $w$ is denoted $w[i]$, i.e. $w = w[0]w[1] \cdots$. Further, we define the ith suffix of $w$ as $w_i = w[i]w[i+1] \cdots$. The semantics of a formula on $w$ is then defined inductively as follows:*

$$\begin{aligned}
w &\models a & &\iff a \in w[0] \\
w &\models \neg a & &\iff a \notin w[0] \\
w &\models \varphi \wedge \psi & &\iff w \models \varphi \text{ and } w \models \psi \\
w &\models \varphi \vee \psi & &\iff w \models \varphi \text{ or } w \models \psi \\
w &\models \mathbf{F}\varphi & &\iff \exists k \in \mathbb{N} : w_k \models \varphi \\
w &\models \mathbf{G}\varphi & &\iff \forall k \in \mathbb{N} : w_k \models \varphi
\end{aligned}$$

We define a symbolic one-step unfolding $\mathfrak{U}$ of a formula inductively by the following rules, where the symbol $\mathbf{X}$ intuitively corresponds to the meaning of the standard next operator.

$$\begin{aligned}
\mathfrak{U}(a) &= a \\
\mathfrak{U}(\neg a) &= \neg a \\
\mathfrak{U}(\varphi \wedge \psi) &= \mathfrak{U}(\varphi) \wedge \mathfrak{U}(\psi) \\
\mathfrak{U}(\varphi \vee \psi) &= \mathfrak{U}(\varphi) \vee \mathfrak{U}(\psi) \\
\mathfrak{U}(\mathbf{F}\varphi) &= \mathfrak{U}(\varphi) \vee \mathbf{XF}\varphi \\
\mathfrak{U}(\mathbf{G}\varphi) &= \mathfrak{U}(\varphi) \wedge \mathbf{XG}\varphi
\end{aligned}$$

*Example 3.* Consider $\varphi = \mathbf{F}a \wedge \mathbf{GF}b$. Then $\mathfrak{U}(\varphi) = (a \vee \mathbf{XF}a) \wedge (b \vee \mathbf{XF}b) \wedge \mathbf{XGF}b$.

## 3 Deterministic Automaton for the (F,G)-fragment

Let $\varphi$ be an arbitrary but fixed formula. In the following, we construct a deterministic finite $\omega$-automaton that recognizes the words satisfying $\varphi$. The definition of the acceptance condition and its variants follow in the subsequent sections. We start with a construction of the state space. The idea is that a state corresponds to a formula that needs to be satisfied when coming into this state. After evaluating the formulae on the propositions currently read, the next state will be given by what remains in the one-step unfold of the formula. E.g. for Example 3 and reading $a$, the successor state needs to satisfy $\mathbf{F}b \wedge \mathbf{GF}b$.

In the classical syntactic model constructions, the states are usually given by sets of subformulae of $\varphi$. This corresponds to the conjunction of these subformulae. The main difference in our approach is the use of both conjunctions and also disjunctions that allow us to dispose of non-determinism in the corresponding transition function. In order to formalize this, we need some notation.

Let $\mathbb{F}$ and $\mathbb{G}$ denote the set of all subformulae of $\varphi$ of the form $\mathbf{F}\psi$ and $\mathbf{G}\psi$, respectively. Further, all temporal subformulae are denoted by a shorthand $\mathbb{T} := \mathbb{F} \cup \mathbb{G}$. Finally, for a set of formulae $\Psi$, we denote $\mathbf{X}\Psi := \{\mathbf{X}\psi \mid \psi \in \Psi\}$.

We denote the *closure* of $\varphi$ by $\mathbb{C}(\varphi) := Ap \cup \{\neg a \mid a \in Ap\} \cup \mathbf{X}\mathbb{T}$. Then $\mathfrak{U}(\varphi)$ is a positive Boolean combination over $\mathbb{C}(\varphi)$. By states($\varphi$) we denote the set $2^{2^{\mathbb{C}(\varphi)}}$. Each element of states($\varphi$) is a positive Boolean function over $\mathbb{C}(\varphi)$ and we often use a positive Boolean formula as its representative. For instance, the definition of $\mathfrak{U}$ is clearly independent of the choice of representative, hence we abuse the notation and apply $\mathfrak{U}$ to elements of states($\varphi$). Note that $|\text{states}(\varphi)| \in \mathcal{O}(2^{2^{|\varphi|}})$ where $|\varphi|$ denotes the length of $\varphi$.

Our state space has two components. Beside the logical component, we also keep track of one-step history of the word read. We usually use letters $\psi, \chi$ when speaking about the former component and $\alpha, \beta$ for the latter one.

**Definition 4.** *Given a formula $\varphi$, we define $\mathcal{A}(\varphi) = (Q, i, \delta)$ to be a deterministic finite automaton over $\Sigma = 2^{Ap}$ given by*

- *the set of states $Q = \{i\} \cup \left(\text{states}(\varphi) \times 2^{Ap}\right)$*
- *the initial state $i$;*
- *the transition function*

$$\delta = \{(i, \alpha, \langle \mathfrak{U}(\varphi), \alpha \rangle) \mid \alpha \in \Sigma\} \cup \{(\langle \psi, \alpha \rangle, \beta, \langle \text{succ}(\psi, \alpha), \beta \rangle) \mid \langle \psi, \alpha \rangle \in Q, \beta \in \Sigma\}$$

*where $\text{succ}(\psi, \alpha) = \mathfrak{U}(\text{next}(\psi[\alpha \mapsto \mathbf{tt}, Ap \setminus \alpha \mapsto \mathbf{ff}])$ where $\text{next}(\psi')$ removes $\mathbf{X}$'s from $\psi'$ and $\psi[T \mapsto \mathbf{tt}, F \mapsto \mathbf{ff}]$ denotes the equivalence class of formulae where in $\psi$ we substitute $\mathbf{tt}$ for all elements of $T$ and $\mathbf{ff}$ for all elements of $F$.*

Intuitively, a state $\langle \psi, \alpha \rangle$ of $Q$ corresponds to the situation where $\psi$ needs to be satisfied and $\alpha$ is being read.

*Example 5.* The automaton for **F**$a$ with $Ap = \{a\}$ is depicted in the following figure. The automaton is obviously unnecessarily large, one can expect to merge e.g. the two states bearing the requirement **tt** as the proposition $a$ is irrelevant for satisfaction of **tt** that does not even contain it. For the sake of simplicity, we leave all possible combinations here and comment on this in Section 8.

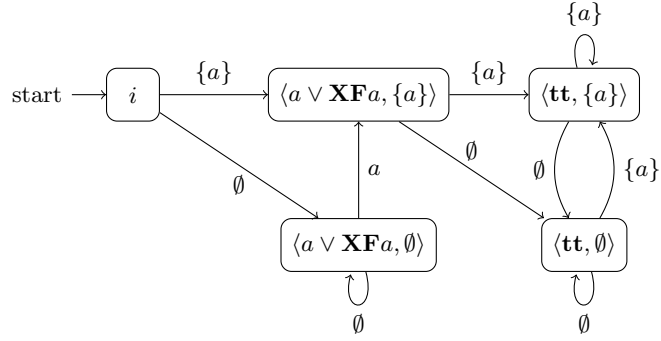

The reader might be surprised or even annoyed by the fact that the logical structure of the state space is not sufficient to keep enough information to decide whether a run $\rho$ is accepting. In order to ensure this, we remember one-step history in the state. Why is that? Consider $\varphi = \mathbf{GF}(a \wedge \mathbf{F}b)$. Its unfold is then

$$\mathbf{XGF}(a \wedge \mathbf{F}b) \wedge \Big(\mathbf{XF}(a \wedge \mathbf{F}b) \vee \big(a \wedge (b \vee \mathbf{XF}b)\big)\Big) \qquad (*)$$

Then moving under $\{a\}$ results into the requirement $\mathbf{GF}(a \wedge \mathbf{F}b) \wedge \big(\mathbf{F}(a \wedge \mathbf{F}b) \vee \mathbf{F}b\big)$ for the next step where the alternative of pure **F**$b$ signals progress made by not having to wait for an $a$. Nevertheless, the unfold of this formula is propositionally equivalent to $(*)$. This is indeed correct as the two formulae are temporally equivalent (i.e. in LTL semantics). Thus, the information about the read $a$ is not kept in the state and the information about this partial progress is lost! And now the next step under both $\{b\}$ and $\emptyset$ again lead to the same requirement $\mathbf{GF}(a \wedge \mathbf{F}b) \wedge \mathbf{F}(a \wedge \mathbf{F}b)$. Therefore, there is no information that if $b$ is read, then it can be matched with the previous $a$ and we already have one satisfaction of (infinitely many required satisfactions of) $\mathbf{F}(a \wedge \mathbf{F}b)$ compared to reading $\emptyset$. Hence, the runs on $(\{a\}\{b\})^\omega$ and $(\{a\}\emptyset)^\omega$ are the same while the former should be accepting and the latter rejecting. However, this can be fixed by remembering the one-step history and using the acceptance condition defined in the following section.

## 4 Muller Acceptance Condition

In this section, we introduce a Muller acceptance condition. In general, the number of sets in a Muller condition can be exponentially larger than the size of the automaton. Therefore, we investigate the particular structure of the condition. In the next section, we provide a much more compact whilst still useful description of the condition. Before giving the formal definition, let us show an example.

*Example 6.* Let $\varphi = \mathbf{F}(\mathbf{G}a \vee \mathbf{G}b)$. The corresponding automaton is depicted below, for clarity, we omit the initial state. Observe that the formula stays the same and the only part that changes is the letter currently read that we remember in the state. The reason why is that $\varphi$ can neither fail in finite time (there is always time to fulfill it), nor can be partially satisfied (no progress counts in this formula, only the infinite suffix). However, at some finite time the argument of $\mathbf{F}$ needs to be satisfied. Although we cannot know when and whether due to $\mathbf{G}a$ or $\mathbf{G}b$, we know it is due to one of these (or both) happening. Thus we may shift the non-determinism to the acceptance condition, which says here: accept if the states where $a$ holds are ultimately never left, or the same happens for $b$. The commitment to e.g. ultimately satisfying $\mathbf{G}a$ can then be proved by checking that all infinitely often visited states read $a$.

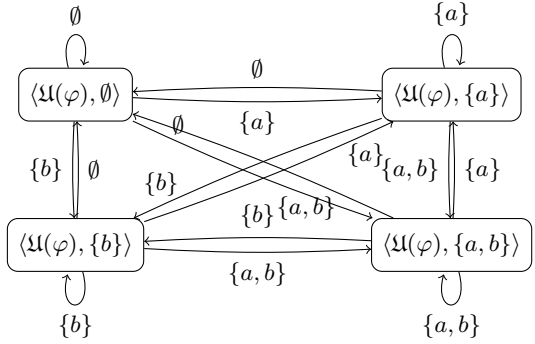

We now formalize this idea. Let $\varphi$ be a formula and $\mathcal{A}(\varphi) = (Q, i, \delta)$ its corresponding automaton. Consider a formula $\chi$ as a Boolean function over elements of $\mathbb{C}(\varphi)$. For sets $T, F \subseteq \mathbb{C}(\varphi)$, let $\chi[T \mapsto \mathbf{tt}, F \mapsto \mathbf{ff}]$ denote the formula where $\mathbf{tt}$ is substituted for elements of $T$, and $\mathbf{ff}$ for $F$. As elements of $\mathbb{C}(\varphi)$ are considered to be atomic expressions here, the substitution is only done on the propositional level and does not go through the modality, e.g. $(a \vee \mathbf{XG}a)[a \to \mathbf{ff}] = \mathbf{ff} \vee \mathbf{XG}a$, which is equivalent to $\mathbf{XG}a$ in the propositional semantics.

Further, for a formula $\chi$ and $\alpha \in \Sigma$ and $I \subseteq \mathbb{T}$, we put $I \models_\alpha \chi$ to denote that

$$\chi[\alpha \cup I \mapsto \mathbf{tt}, Ap \setminus \alpha \mapsto \mathbf{ff}]$$

is equivalent to $\mathbf{tt}$ in the propositional semantics. We use this notation to describe that we rely on a commitment to satisfy all formulae of $I$.

**Definition 7 (Muller acceptance).** *A set $M \subseteq Q$ is Muller accepting for a set $I \subseteq \mathbb{T}$ if the following is satisfied:*

1. *for each $(\chi, \alpha) \in M$, we have $\mathbf{X}I \models_\alpha \chi$,*
2. *for each $\mathbf{F}\psi \in I$ there is $(\chi, \alpha) \in M$ with $I \models_\alpha \psi$,*
3. *for each $\mathbf{G}\psi \in I$ and for each $(\chi, \alpha) \in M$ we have $I \models_\alpha \psi$.*

*A set $F \subseteq Q$ is Muller accepting (for $\varphi$) if it is Muller accepting for some $I \subseteq \mathbb{T}$.*

The first condition ensures that the commitment to formulae in $I$ being ultimately satisfied infinitely often is enough to satisfy the requirements. The second one guarantees that each **F**-formula is unfolded only finitely often and then satisfied, while the third one guarantees that **G**-formulae indeed ultimately hold. Note that it may be impossible to see the satisfaction of a formula directly and one must rely on further promises, formulae of smaller size. In the end, promising the atomic proposition is not necessary and is proven directly from the second component of the state space.

### 4.1 Correctness

Given a formula $\varphi$, we have defined a Muller automaton $\mathcal{A}(\varphi)$ and we let the acceptance condition $\mathcal{M}(\varphi) = \{M_1, \ldots, M_k\}$ be given by all the Muller accepting sets $M_i$ for $\varphi$. Every word $w : \mathbb{N} \to 2^{Ap}$ induces a run $\rho = \mathcal{A}(\varphi)(w) : \mathbb{N} \to Q$ starting in $i$ and following $\delta$. The run is thus accepting and the word is accepted if the set of states visited infinitely often $\mathrm{Inf}(\rho)$ is Muller accepting for $\varphi$. Vice versa, a run $\rho = i(\chi_1, \alpha_1)(\chi_2, \alpha_2) \cdots$ induces a word $Ap(\rho) = \alpha_1 \alpha_2 \cdots$. We now prove that this acceptance condition is sound and complete.

**Theorem 8.** *Let $\varphi$ be a formula and $w$ a word. Then $w$ is accepted by the deterministic automaton $\mathcal{A}(\varphi)$ with the Muller condition $\mathcal{M}(\varphi)$ if and only if $w \models \varphi$.*

We start by proving that the first component of the state space takes care of all progress or failure in finite time.

**Proposition 9 (Local (finitary) correctness).** *Let $w$ be a word and $\mathcal{A}(\varphi)(w) = i(\chi_0, \alpha_0)(\chi_1, \alpha_1) \cdots$ the corresponding run. Then for all $n \in \mathbb{N}$, we have $w \models \varphi$ if and only if $w_n \models \chi_n$.*

*Proof (Sketch).* The one-step unfold produces a temporally equivalent (w.r.t. LTL satisfaction) formula. The unfold is a Boolean function over atomic propositions and elements of $\mathbf{X}\mathbb{T}$. Therefore, this unfold is satisfied if and only if the next state satisfies $\mathrm{next}(\psi)$ where $\psi$ is the result of partial application of the Boolean function to the currently read letter of the word. We conclude by induction. □

Further, each occurrence of satisfaction of **F** must happen in finite time. As a consequence, a run with $\chi_i \not\equiv \mathbf{ff}$ is rejecting if and only if satisfaction of some $\mathbf{F}\psi$ is always postponed.

**Proposition 10 (Completeness).** *If $w \models \varphi$ then $\mathrm{Inf}(\mathcal{A}(\varphi)(w))$ is a Muller accepting set.*

*Proof.* Let us show that $M := \mathrm{Inf}(\mathcal{A}(\varphi)(w))$ is Muller accepting for

$$I := \{\psi \in \mathbb{F} \mid w \models \mathbf{G}\psi\} \cup \{\psi \in \mathbb{G} \mid w \models \mathbf{F}\psi\}$$

As a technical device we use the following. For every finite Boolean combination $\psi$ of elements of the closure $\mathbb{C}$, there are only finitely many options to satisfy it, each corresponding to a subset of $\mathbb{C}$. Therefore, if $w_i \models \psi$ for infinitely many $i \in \mathbb{N}$ then at least one of the options has to recur. More precisely, for some subset $\alpha \subseteq Ap$ there are infinitely many $i \in \mathbb{N}$ with $w_i \models \psi \cup \alpha \cup \{\neg a \mid a \in Ap \setminus \alpha\}$. For each such $\alpha$ we pick one subset $I_{\chi,\alpha} \subseteq \mathbb{T}$ such that for infinitely many $i$, after reading $w^i = w[0] \cdots w[i]$ we are in state $(\chi, \alpha)$ and $w_i \models \psi \cup \mathbf{X} I_{\chi,\alpha}$, and $I_{\chi,\alpha} \models_\alpha \psi$. We say that we have a *recurring set $I_{\chi,\alpha}$ modelling $\psi$ (for a state $(\chi, \alpha)$)*. Obviously, the recurring sets for all states are included in $I$, i.e. $I_{\chi,\alpha} \subseteq I$ for every $(\chi, \alpha) \in Q$.

Let us now proceed with proving the three conditions of Definition 7 for $M$ and $I$.

Condition 1. Let $(\chi, \alpha) \in M$. Since $w \models \varphi$, by Proposition 9 $w_i \models \chi$ whenever we enter $(\chi, \alpha)$ after reading $w^i$, which happens for infinitely many $i \in \mathbb{N}$. Hence we have a recurring set $I_{\chi,\alpha}$ modelling $\chi$. Since $I_{\chi,\alpha} \models_\alpha \chi$, we get also $I \models_\alpha \chi$ by $I_{\chi,\alpha} \subseteq I$.

Condition 2. Let $\mathbf{F}\psi \in I$, then $w \models \mathbf{GF}\psi$. Since there are finitely many states, there is $(\chi, \alpha) \in M$ for which after infinitely many entrances by $w^i$ it holds $w_i \models \psi$ by Proposition 9, hence we have a recurring set $I_{\chi,\alpha}$ modelling $\psi$ and conclude as above.

Condition 3. Let $\mathbf{G}\psi \in I$, then $w \models \mathbf{FG}\psi$. Hence for every $(\chi, \alpha) \in M$ infinitely many $w^i$ leading to $(\chi, \alpha)$ satisfy $w_i \models \psi$ by Proposition 9, hence we have a recurring set $I_{\chi,\alpha}$ modelling $\psi$ and conclude as above. □

Before proving the opposite direction of the theorem, we provide a property of Muller accepting sets opposite to the previous proposition.

**Lemma 11.** *Let $\rho$ be a run. If $\mathrm{Inf}(\rho)$ is Muller accepting for $I$ then $Ap(\rho) \models \mathbf{G}\psi$ for each $\psi \in I \cap \mathbb{F}$ and $Ap(\rho) \models \mathbf{F}\psi$ for each $\psi \in I \cap \mathbb{G}$.*

*Proof.* Denote $w = Ap(\rho)$. Let us first assume $\psi \in I \cap \mathbb{F}$ and $w_j \not\models \psi$ for all $j \geq i \in \mathbb{N}$. Since $\psi \in I \cap \mathbb{F}$, for infinitely many $j$, $\rho$ passes through some $(\chi, \alpha) \in \mathrm{Inf}(\rho)$ for which $I \models_\alpha \psi$. Hence, there is $\psi_1 \in I$ which is a subformula of $\psi$ such that for infinitely many $i$, $w_i \not\models \psi_1$. If $\psi_1 \in \mathbb{F}$, we proceed as above; similarly for $\psi_1 \in \mathbb{G}$. Since we always get a smaller subformula, at some point we obtain either $\psi_n = \mathbf{F}\beta$ or $\psi_n = \mathbf{G}\beta$ with $\beta$ a Boolean combination over $Ap$ and we get a contradiction with the second or the third point of Definition 7, respectively. □

In other words, if we have a Muller accepting set for $I$ then all elements of $I$ hold true in $w_i$ for almost all $i$.

**Proposition 12 (Soundness).** *If $\mathrm{Inf}(\mathcal{A}(\varphi)(w))$ is a Muller accepting set then $w \models \varphi$.*

*Proof.* Let $M := \mathrm{Inf}(\mathcal{A}(\varphi)(w))$ be a Muller accepting set for some $I$. There is $i \in \mathbb{N}$ such that after reading $w^i$ we come to $(\chi, \alpha)$ and stay in $\mathrm{Inf}(\mathcal{A}(\varphi)(w))$ from now on and, moreover, $w_i \models \psi$ for all $\psi \in I$ by Lemma 11. For a contradiction, let $w \not\models \varphi$. By Proposition 9 we thus get $w_i \not\models \chi$. By the first condition of Definition 7, we get $I \models_\alpha \chi$. Therefore, there is $\psi \in I$ such that $w_i \not\models \psi$, a contradiction. □

## 5  Generalized Rabin Condition

In this section, we investigate the structure of the previously defined Muller condition and propose a new type of acceptance condition that compactly, yet reasonably explicitly captures the accepting sets.

Let us first consider a fixed $I \subseteq \mathbb{T}$ and examine all Muller accepting sets for $I$. The first condition of Definition 7 requires not to leave the set of states $\{(\chi, \alpha) \mid I \models_\alpha \chi)\}$. Similarly, the third condition is a conjunction of $|I \cap \mathbb{G}|$ conditions not to leave sets $\{(\chi, \alpha) \mid I \models_\alpha \psi\}$ for each $\mathbf{G}\psi \in I$. Both conditions thus together require that certain set (complement of the intersection of the above sets) is visited only finitely often. On the other hand, the second condition requires to visit certain sets infinitely often. Indeed, for each $\mathbf{F}\psi$ the set $\{(\chi, \alpha) \mid I \models_\alpha \psi\}$ must be visited infinitely often.

Furthermore, a set is accepting if the conditions above hold for *some* set $I$. Hence, the acceptance condition can now be expressed as a positive Boolean combination over Rabin pairs in a similar way as the standard Rabin condition is a disjunction of Rabin pairs.

*Example 13.* Let us consider the (strong) fairness constraint $\varphi = \mathbf{FG}a \vee \mathbf{GF}b$. Since each atomic proposition has both $\mathbf{F}$ and $\mathbf{G}$ as ancestors in the syntactic tree, it is easy to see that there is only one reachable element of states$(\varphi)$ and the state space of $\mathcal{A}$ is $\{i\} \cup 2^{\{a,b\}}$, i.e. of size $1 + 2^2 = 5$. Furthermore, the syntactic tree of $\mathfrak{U}(\varphi) = \mathbf{XFG}a \vee (\mathbf{XG}a \wedge a) \vee (\mathbf{XGF}b \wedge (\mathbf{XF}b \vee b))$ immediately determines possible sets $I$. These either contain $\mathbf{G}a$ (possibly with also $\mathbf{FG}a$ or some other elements) or $\mathbf{GF}b, \mathbf{F}b$. The first option generates the requirement to visit states with $\neg a$ only finitely often, the second one to visit $b$ infinitely often. Thus the condition can be written as

$$(\{q \mid q \models \neg a\}, Q) \vee (\emptyset, \{q \mid q \models b\})$$

and is in fact a Rabin acceptance condition.

We formalize this new type of acceptance condition as follows.

**Definition 14 (Generalized Rabin Automaton).** *A* generalized Rabin automaton *is a (deterministic) $\omega$-automaton $\mathcal{A} = (Q, i, \delta)$ over some alphabet $\Sigma$, where $Q$ is a set of states, $i$ is the initial state, $\delta : Q \times \Sigma \to Q$ is a transition function, together with a* generalized Rabin condition $\mathcal{GR} \in \mathcal{B}^+(2^Q \times 2^Q)$. *A run $\rho$ of $\mathcal{A}$ is accepting if $\mathrm{Inf}(\rho) \models \mathcal{GR}$, which is defined inductively as follows:*

$$\begin{aligned}
\mathrm{Inf}(\rho) \models \varphi \wedge \psi &\iff \mathrm{Inf}(\rho) \models \varphi \text{ and } \mathrm{Inf}(\rho) \models \psi \\
\mathrm{Inf}(\rho) \models \varphi \vee \psi &\iff \mathrm{Inf}(\rho) \models \varphi \text{ or } \mathrm{Inf}(\rho) \models \psi \\
\mathrm{Inf}(\rho) \models (F, I) &\iff F \cap \mathrm{Inf}(\rho) = \emptyset \text{ and } I \cap \mathrm{Inf}(\rho) \neq \emptyset
\end{aligned}$$

The generalized Rabin condition corresponding to the previously defined Muller condition $\mathcal{M}$ can now be formalized as follows.

**Definition 15 (Generalized Rabin Acceptance).** *Let $\varphi$ be a formula. The generalized Rabin condition $\mathcal{GR}(\varphi)$ is*

$$\bigvee_{I \subseteq \mathbb{T}} \left( \left( \{(\chi, \alpha) \mid I \not\models_\alpha \chi \wedge \bigwedge_{\mathbf{G}\psi \in I} \psi\}, Q \right) \wedge \bigwedge_{\mathbf{F}\omega \in I} \left( \emptyset, \{(\chi, \alpha) \mid I \models_\alpha \omega\} \right) \right)$$

By the argumentation above, we get the equivalence of the Muller and the generalized Rabin conditions for $\varphi$ and thus the following.

**Proposition 16.** *Let $\varphi$ be a formula and $w$ a word. Then $w$ is accepted by the deterministic automaton $\mathcal{A}(\varphi)$ with the generalized Rabin condition $\mathcal{GR}(\varphi)$ if and only if $w \models \varphi$.*

*Example 17.* Let us consider a conjunction of two (strong) fairness constraints $\varphi = (\mathbf{FG}a \vee \mathbf{GF}b) \wedge (\mathbf{FG}c \vee \mathbf{GF}d)$. Since each atomic proposition is wrapped in either $\mathbf{FG}$ or $\mathbf{GF}$, there is again only one relevant element of states($\varphi$) and the state space of $\mathcal{A}$ is $\{i\} \cup 2^{\{a,b,c,d\}}$, i.e. of size $1 + 2^4 = 17$. From the previous example, we already know the disjunctions correspond to $(\neg a, Q) \vee (\emptyset, b)$ and $(\neg c, Q) \vee (\emptyset, d)$. Thus for the whole conjunction, we get a generalized Rabin condition

$$\left( (\neg a, Q) \vee (\emptyset, b) \right) \wedge \left( (\neg c, Q) \vee (\emptyset, d) \right)$$

## 6 Rabin Condition

In this section, we briefly describe how to obtain a Rabin automaton from $\mathcal{A}(\varphi)$ and the generalized Rabin condition $\mathcal{GR}(\varphi)$ of Definition 15. For a fixed $I$, the whole conjunction of Definition 15 corresponds to the intersection of automata with different Rabin conditions. In order to obtain the intersection, one has first to construct the product of the automata, which in this case is still the original automaton with the state space $Q$, as they are all the same. Further, satisfying

$$(G, Q) \wedge \bigwedge_{f \in \mathcal{F} := I \cap \mathbb{F}} (\emptyset, F_f)$$

amounts to visiting $G$ only finitely often and each $F_f$ infinitely often. To check the latter (for a non-empty conjunction), it is sufficient to multiply the state space by $\mathcal{F}$ with the standard trick that we leave the $f$th copy once we visit $F_f$ and immediately go to the next copy. The resulting Rabin pair is thus

$$\left( G \times \mathcal{F}, F_{\bar{f}} \times \{\bar{f}\} \right)$$

for an arbitrary fixed $\bar{f} \in \mathcal{F}$.

As for the disjunction, Rabin condition is closed under it as it simply takes the union of the pairs when the two automata have the same state space. In our case, one can multiply the state space of each disjunct corresponding to $I$ by all

$J \cap \mathbb{F}$ for each $J \in 2^{\mathbb{T}} \setminus \{I\}$ to get the same state space for all of them. We thus get a bound for the state space

$$\prod_{I \subseteq \mathbb{T}} |I \cap \mathbb{F}| \cdot |Q|$$

*Example 18.* The construction of Definition 15 for the two fairness constraints Example 17 yields

$$(\neg a \vee \neg c, Q) \vee (\neg a, d) \vee (\neg c, b) \vee \big((\emptyset, b) \wedge (\emptyset, d)\big)$$

where we omitted all pairs $(F, I)$ for which we already have a pair $(F', I')$ with $F \subseteq F'$ and $I \supseteq I'$. One can eliminate the conjunction as described above at the cost of multiplying the state space by two. The corresponding Rabin automaton thus has $2 \cdot 1 \cdot |\{i\} \cup 2^{Ap}| = 34$ states. (Of course, for instance the initial state need not be duplicated, but for the sake of simplicity of the construction we avoid any optimizations.)

For a conjunction of three conditions, $\varphi = (\mathbf{FG}a \vee \mathbf{GF}b) \wedge (\mathbf{FG}c \vee \mathbf{GF}d) \wedge (\mathbf{FG}e \vee \mathbf{GF}f)$, the right components of the Rabin pairs correspond to $\mathbf{tt}, b, d, f, b \wedge d, b \wedge f, d \wedge f, b \wedge d \wedge f$. The multiplication factor to obtain a Rabin automaton is thus $2 \cdot 2 \cdot 2 \cdot 3 = 24$ and the state space is of the size $24 \cdot 1 \cdot (1 + 2^6) = 1560$.

## 7  Complexity

In this section, we summarize the theoretical complexity bounds we have obtained.

The traditional approach first translates the formula $\varphi$ of length $n$ into a non-deterministic automaton of size $\mathcal{O}(2^n)$. Then the determinization follows. The construction of Safra has the complexity $m^{\mathcal{O}(m)}$ where $m$ is the size of the input automaton [Saf88]. This is in general optimal. The overall complexity is thus

$$2^{n \cdot \mathcal{O}(2^n)} = 2^{\mathcal{O}(2^{n + \log n})}$$

The recent lower bound for the whole LTL is $2^{2^{\Omega(n)}}$ [KR10]. However, to be more precise, the example is of size less than $2^{\mathcal{O}(2^n)}$. Hence, there is a small gap. To the authors' best knowledge, there is no better upper bound when restricting to automata arising from LTL formulae or from the full $(\mathbf{F},\mathbf{G})$-fragment. (There are results on smaller fragments [AT04] though.) We tighten this gap slightly as shown below. Further, note that the number of Rabin pairs is $\mathcal{O}(m) = \mathcal{O}(2^n)$.

Our construction first produces a Muller automaton of size

$$\mathcal{O}(2^{2^{|\mathbb{T}|}} \cdot 2^{|Ap|}) = \mathcal{O}(2^{2^n + n}) \subseteq 2^{\mathcal{O}(2^n)}$$

which is strictly less than in the traditional approach. Moreover, as already discussed in Example 13, one can consider an "infinitary" fragment where every atomic proposition has in the syntactic tree both $\mathbf{F}$ and $\mathbf{G}$ as some ancestors. In this fragment, the state space of the Muller/generalized Rabin automaton is simply $2^{Ap}$ (when omitting the initial state) as for all $\alpha \subseteq Ap$, we have $succ(\varphi, \alpha) = \varphi$. This is useful, since for e.g. fairness constraints our procedure yields exponentially smaller automaton.

Although the size of the Muller acceptance condition can be potentially exponentially larger than the state space, we have shown it can be compactly written as a disjunction of up to $2^n$ of conjunctions each of size at most $n$.

Moreover, using the intersection procedure we obtain a Rabin automaton with the upper bound on the state space

$$|\mathbb{F}|^{2^{|\mathbb{T}|}} \cdot |Q| \in n^{2^n} \cdot 2^{\mathcal{O}(2^n)} = 2^{\mathcal{O}(\log n \cdot 2^n)} = 2^{\mathcal{O}(2^{n+\log \log n})} \subsetneq 2^{\mathcal{O}(2^{n+\log n})}$$

thus slightly improving the upper bound. Further, each conjunction is transformed into one pair, we are thus left with at most $2^{|\mathbb{T}|} \in \mathcal{O}(2^n)$ Rabin pairs.

## 8 Experimental Results and Evaluation

We have implemented the construction of the state space of $\mathcal{A}(\varphi)$ described above. Further, Definition 15 then provides a way to compute the multiplication factor needed in order to get the Rabin automaton. We compare the sizes of this generalized Rabin automaton and Rabin automaton with the Rabin automaton produced by ltl2dstar. Ltl2dstar first calls an external translator from LTL to non-deterministic Büchi automata. In our experiments, it is LTL2BA [GO01] recommended by the authors of ltl2dstar. Then it performs Safra's determinization. Ltl2dstar implements several optimizations of Safra's construction. The optimizations shrink the state space by factor of 5 (saving 79.7% on average on the formulae considered here) to 10 (89.7% on random formulae) [KB06]. Our implementation does not perform any ad hoc optimization, since we want to evaluate whether the basic idea of the Safraless construction is already competitive. The only optimizations done are the following.

- Only the reachable part of the state space is generated.
- Only atomic propositions relevant in each state are considered. In a state $(\chi, \alpha)$, $a$ is not relevant if $\chi[a \mapsto \mathbf{tt}] \equiv \chi[a \mapsto \mathbf{ff}]$, i.e. if for every valuation, $\chi$ has the same value no matter which value $a$ takes. For instance, let $Ap = \{a, b\}$ and consider $\chi = \mathfrak{U}(\mathbf{F}a) = \mathbf{F}a \vee a$. Then instead of having four copies (for $\emptyset, \{a\}, \{b\}, \{a, b\}$), there are only two for the *sets* of valuations $\{\emptyset, \{b\}\}$ and $\{\{a\}, \{a, b\}\}$. For its successor $\mathbf{tt}$, we only have one copy standing for the whole set $\{\emptyset, \{a\}, \{b\}, \{a, b\}\}$.
- Definition 15 takes a disjunction over $I \in 2^{\mathbb{T}}$. If $I \subseteq I'$ but the set of states $(\chi, \alpha)$ with $I \models_\alpha \chi$ and $I' \models_\alpha \chi$ are the same, it is enough to consider the disjunct for $I$ only. E.g. for $\mathfrak{U}(\mathbf{G}(\mathbf{F}a \vee \mathbf{F}b))$, we only consider $I$ either $\{\mathbf{G}(\mathbf{F}a \vee \mathbf{F}b), \mathbf{F}a\}$ or $\{\mathbf{G}(\mathbf{F}a \vee \mathbf{F}b), \mathbf{F}b\}$, but not their union.

  This is an instance of a more general simplification. For a conjunction of pairs $(F_1, I_1) \wedge (F_2, I_2)$ with $I_1 \subseteq I_2$, there is a single equivalent condition $(F_1 \cup F_2, I_1)$.

Table 1 shows the results on formulae from BEEM (BEnchmarks for Explicit Model checkers)[Pel07] and formulae from [SB00] on which ltl2dstar was originally tested [KB06]. In both cases, we only take formulae of the (**F**,**G**)-fragment. In the first case this is 11 out of 20, in the second 12 out of 28. There is a slight overlap between the two sets. Further, we add conjunctions of strong fairness conditions and a few other formulae. For each formula $\varphi$, we give the number $|\text{states}(\varphi)|$ of distinct states w.r.t. the first (logical) component. The overall number of states of the Muller or generalized Rabin automaton follows. The respective runtimes are not listed as they were less than a second for all listed formulae, with the exception of the fifth formula from the bottom where it needed 3 minutes (here ltl2dstar needed more than one day to compute the Rabin automaton). In the column $\mathcal{GR}$-factor, we describe the complexity of the generalized Rabin condition, i.e. the number of copies of the state space that are created to obtain an equivalent Rabin automaton, whose size is thus bounded from above by the column Rabin. The last column states the size of the state space of the Rabin automaton generated by ltl2dstar using LTL2BA.

**Table 1.** Experimental comparison to ltl2dstar on formulae of [Pel07], [SB00], fairness constraints and some other examples of formulae of the "infinitary" fragment

| Formula | states | Muller/GR | $\mathcal{GR}$-factor | Rabin | ltl2dstar |
|---|---|---|---|---|---|
| $\mathbf{G}(a \vee \mathbf{F}b)$ | 2 | 5 | 1 | 5 | 4 |
| $\mathbf{FG}a \vee \mathbf{FG}b \vee \mathbf{GF}c$ | 1 | 9 | 1 | 9 | 36 |
| $\mathbf{F}(a \vee b)$ | 2 | 4 | 1 | 4 | 2 |
| $\mathbf{GF}(a \vee b)$ | 1 | 3 | 1 | 3 | 4 |
| $\mathbf{G}(a \vee b \vee c)$ | 2 | 4 | 1 | 4 | 3 |
| $\mathbf{G}(a \vee \mathbf{F}b)$ | 2 | 5 | 1 | 5 | 4 |
| $\mathbf{G}(a \vee \mathbf{F}(b \vee c))$ | 2 | 5 | 1 | 5 | 4 |
| $\mathbf{F}a \vee \mathbf{G}b$ | 3 | 7 | 1 | 7 | 5 |
| $\mathbf{G}(a \vee \mathbf{F}(b \wedge c))$ | 2 | 5 | 1 | 5 | 4 |
| $(\mathbf{FG}a \vee \mathbf{GF}b)$ | 1 | 5 | 1 | 5 | 12 |
| $\mathbf{GF}(a \vee b) \wedge \mathbf{GF}(b \vee c)$ | 1 | 5 | 2 | 10 | 12 |
| $(\mathbf{FF}a \wedge \mathbf{G}\neg a) \vee (\mathbf{GG}\neg a \wedge \mathbf{F}a)$ | 2 | 4 | 1 | 4 | 1 |
| $(\mathbf{GF}a) \wedge \mathbf{FG}b$ | 1 | 5 | 1 | 5 | 7 |
| $(\mathbf{GF}a \wedge \mathbf{FG}b) \vee (\mathbf{FG}\neg a \wedge \neg b)$ | 1 | 5 | 1 | 5 | 14 |
| $\mathbf{FG}a \wedge \mathbf{GF}a$ | 1 | 3 | 1 | 3 | 3 |
| $\mathbf{G}(\mathbf{F}a \wedge \mathbf{F}b)$ | 1 | 5 | 2 | 10 | 5 |
| $\mathbf{F}a \wedge \mathbf{F}b$ | 4 | 8 | 1 | 8 | 4 |
| $(\mathbf{G}(b \vee \mathbf{GF}a) \wedge \mathbf{G}(c \vee \mathbf{GF}\neg a)) \vee \mathbf{G}b \vee \mathbf{G}c$ | 4 | 18 | 2 | 36 | 26 |
| $(\mathbf{G}(b \vee \mathbf{FG}a) \wedge \mathbf{G}(c \vee \mathbf{FG}\neg a)) \vee \mathbf{G}b \vee \mathbf{G}c$ | 4 | 18 | 1 | 18 | 29 |
| $(\mathbf{F}(b \wedge \mathbf{FG}a) \vee \mathbf{F}(c \wedge \mathbf{FG}\neg a)) \wedge \mathbf{F}b \wedge \mathbf{F}c$ | 4 | 18 | 1 | 18 | 8 |
| $(\mathbf{F}(b \wedge \mathbf{GF}a) \vee \mathbf{F}(c \wedge \mathbf{GF}\neg a)) \wedge \mathbf{F}b \wedge \mathbf{F}c$ | 4 | 18 | 1 | 18 | 45 |
| $(\mathbf{FG}a \vee \mathbf{GF}b)$ | 1 | 5 | 1 | 5 | 12 |
| $(\mathbf{FG}a \vee \mathbf{GF}b) \wedge (\mathbf{FG}c \vee \mathbf{GF}d)$ | 1 | 17 | 2 | 34 | 17 527 |
| $\bigwedge_{i=1}^{3}(\mathbf{GF}a_i \to \mathbf{GF}b_i)$ | 1 | 65 | 24 | 1 560 | 1 304 706 |
| $(\bigwedge_{i=1}^{5} \mathbf{GF}a_i) \to \mathbf{GF}b$ | 1 | 65 | 1 | 65 | 972 |
| $\mathbf{GF}(\mathbf{F}a\mathbf{GF}b\mathbf{FG}(a \vee b))$ | 1 | 5 | 1 | 5 | 159 |
| $\mathbf{FG}(\mathbf{F}a \vee \mathbf{GF}b \vee \mathbf{FG}(a \vee b))$ | 1 | 5 | 1 | 5 | 2918 |
| $\mathbf{FG}(\mathbf{F}a \vee \mathbf{GF}b \vee \mathbf{FG}(a \vee b) \vee \mathbf{FG}b)$ | 1 | 5 | 1 | 5 | 4516 |

While the advantages of our approach over the general determinization are clear for the infinitary fragment, there seem to be some drawbacks when "finitary" behaviour is present, i.e. behaviour that can be satisfied or disproved after finitely many steps. The reason and the patch for this are the following. Consider the formula $\mathbf{F}a$ and its automaton from Example 5. Observe that one can easily collapse the automaton to the size of only 2. The problem is that some states such as $\langle a \vee \mathbf{XF}a, \{a\}\rangle$ are only "passed through" and are equivalent to some of their successors, here $\langle \mathbf{tt}, \{a\}\rangle$. However, we may safely perform the following collapse. Whenever two states $(\chi, \alpha), (\chi', \alpha)$ satisfy that $\chi[\alpha \mapsto \mathbf{tt}, Ap \setminus \alpha \mapsto \mathbf{ff}]$ is propositionally equivalent to $\chi'[\alpha \mapsto \mathbf{tt}, Ap \setminus \alpha \mapsto \mathbf{ff}]$ we may safely merge the states as they have the same properties: they are bisimilar with the same set of atomic propositions satisfied. Using these optimizations, e.g. the automaton for $\mathbf{F}a \wedge \mathbf{F}b$ has size 4 as the one produced by ltl2dstar.

Next important observation is that the blow-up from generalized Rabin to Rabin automaton (see the column $\mathcal{GR}$-factor) corresponds to the number of elements of $\mathbb{F}$ that have a descendant or an ancestor in $\mathbb{G}$ and are combined with conjunction. This follows directly from the transformation described in Section 6 and is illustrated in the table.

Thus, we may conclude that our approach is competitive to the determinization approach and for some classes of useful properties such as fairness constraints or generally the infinitary properties it shows significant advantages. Firstly, the state space of the Rabin automaton is noticeably smaller. Secondly, compact generalized Rabin automata tend to be small even for more complex formulae. Thirdly, the state spaces of our automata have a clear structure to be exploited for further possible optimizations, which is more difficult in the case of determinization. In short, the state space is less "messy".

## 9 Discussion on Extensions

Our approach seems to be extensible to the ($\mathbf{X},\mathbf{F},\mathbf{G}$)-fragment. In this setting, instead of remembering the one-step history one needs to remember $n$ last steps (or have a $n$-step look-ahead) in order to deal with formulae such as $\mathbf{GF}(a \wedge \mathbf{X}b)$. Indeed, the acceptance condition requires to visit infinitely often a state provably satisfying $a \wedge \mathbf{X}b$. This can be done by remembering the last $n$ symbols read, where $n$ can be chosen to be the nesting depth of $\mathbf{X}$s. We have not presented this extension mainly for the sake of clarity of the construction.

Further, one could handle the positive ($\mathbf{X},\mathbf{U}$)-fragment, where only atomic propositions may be negated as defined above. These formulae are purely "finitary" and the logical component of the state space is sufficient. Indeed, the automaton simply accepts if and only if $\mathbf{tt}$ is reached and there is no need to check any formulae that we had committed to.

For the ($\mathbf{U},\mathbf{G}$)-fragment or the whole LTL, our approach would need to be significantly enriched as the state space (and last $n$ symbols read) is not sufficient to keep enough information to decide whether a run $\rho$ is accepting only based on $\mathrm{Inf}(\rho)$. Indeed, consider a formula $\varphi = \mathbf{GF}(a \wedge b\mathbf{U}c)$. Then reading $\{a, b\}$ results in the requirement $\mathbf{GF}(a \wedge b\mathbf{U}c) \wedge \big(\mathbf{F}(a \wedge b\mathbf{U}c) \vee (b\mathbf{U}c)\big)$ which is, however, temporally equivalent to $\varphi$ (their unfolds are propositionally equivalent). Thus, runs on $(\{a,b\}\{c\}\emptyset)^\omega$ and $(\{a,b\}\emptyset\{c\})^\omega$ have the same set of infinitely often visited states. Hence, the order of visiting the states matters and one needs the history. However, words such as $(\{a,b\}\{b\}^n\{c\})^\omega$ vs. $(\{b\}^n\{c\})^\omega$ show that more complicated structure is needed than last $n$ letters. The conjecture that this approach is extensible to the whole LTL is left open and considered for future work.

## 10  Conclusions

We have shown a direct translation of the LTL fragment with operators $\mathbf{F}$ and $\mathbf{G}$ to deterministic automata. This translation has several advantages compared to the traditional way that goes via non-deterministic Büchi automata and then performs determinization. First of all, in our opinion it is a lot simpler than the determinization and its various non-trivial optimizations. Secondly, the state space has a clear logical structure. Therefore, any work with the automata or further optimizations seem to be conceptually easier. Moreover, many optimizations are actually done by the logic itself. Indeed, logical equivalence of the formulae helps to shrink the state space with no further effort. In a sense, the logical part of a state contains precisely the information that the semantics of LTL dictates, see Proposition 9. Thirdly, the state space is—according to the experiments—not much bigger even when compared to already optimized determinization. Moreover, very often it is considerably smaller, especially for the "infinitary" formulae; in particular, for fairness conditions. Furthermore, we have also given a very compact deterministic $\omega$-automaton with a small and in our opinion reasonably simple generalized Rabin acceptance condition.

Although we presented a possible direction to extend the approach to the whole LTL, we leave this problem open and will focus on this in future work. Further, since only the obvious optimizations mentioned in Section 8 have been implemented so far, there is space for further performance improvements in this new approach.

### Acknowledgement

Thanks to Andreas Gaiser for pointing out to us that ltl2dstar constructs surprisingly large automata for fairness constraints and the anonymous reviewers for their valuable comments.